\documentclass{sig-alternate-10pt}

\usepackage{algorithm}
\usepackage{algpseudocode}
\usepackage{blindtext}
\usepackage{booktabs}
\usepackage[inline]{enumitem}
\usepackage[T1]{fontenc}
\usepackage{graphicx}
\usepackage[utf8]{inputenc}
\usepackage{multirow}
\usepackage{numprint}
\usepackage[group-separator={,}]{siunitx}
\usepackage{subfigure}
\usepackage{wrapfig}
\usepackage[usenames,dvipsnames]{xcolor}
\usepackage{url}

\usepackage[pdfstartview=FitH, 
  pdfpagelayout=SinglePage,
  bookmarksnumbered=true,
  bookmarksopen=true,
  colorlinks=true,
  citecolor=blue,
  linkcolor=blue,
  urlcolor=blue,
  breaklinks=true]{hyperref}

\newcommand{\parab}[1]{\vspace{0.02in}\noindent{\textbf{#1}}}

\newcommand{\paragraphb}[1]{\vspace{0.02in} \noindent \textbf{#1}}
\newcommand{\cut}[1]{}

\newcommand{\ukm}[1]{\SI{#1}{\kilo\metre}}
\newcommand{\ukmps}[1]{\SI{#1}{\kilo\metre/\second}}

\newcolumntype{L}[1]{>{\raggedright\let\newline\\\arraybackslash\hspace{0pt}}m{#1}}
\newcolumntype{C}[1]{>{\centering\let\newline\\\arraybackslash\hspace{0pt}}m{#1}}
\newcolumntype{R}[1]{>{\raggedleft\let\newline\\\arraybackslash\hspace{0pt}}m{#1}}

\newcommand{\ie}{{\it i.e.,}\xspace}
\newcommand{\eg}{{\it e.g.,}\xspace}
\newcommand{\comment}[1]{}

\begin{document}


\title{Dissecting Latency in the Internet's Fiber Infrastructure}

\author{
Ilker Nadi Bozkurt\footnotemark[1]~~~
Waqar Aqeel\footnotemark[1]~~~
Debopam Bhattacherjee\footnotemark[2]~~~
Balakrishnan Chandrasekaran\footnotemark[3]\\
Philip Brighten Godfrey\footnotemark[4]~~~~
Gregory Laughlin\footnotemark[5]~~~~
Bruce M. Maggs\footnotemark[1]~\footnotemark[6]~~~~
Ankit Singla\footnotemark[2]
~\\
\affaddr{\footnotemark[1]~\,Duke University}~~~~
\affaddr{\footnotemark[2]~\,ETH Z\"{u}rich} ~~~~
\affaddr{\footnotemark[3]~\,Max-Planck-Instit\"{u}t f\"{u}r Informatik}\\
\affaddr{\footnotemark[4]~\,UIUC}~~~~
\affaddr{\footnotemark[5]~\,Yale University}~~~~
\affaddr{\footnotemark[6]~\,Akamai Technologies} 
}

\maketitle

\abstract{The recent publication of the ``InterTubes'' map of long-haul fiber-optic cables in the contiguous United States invites an exciting question: how much faster would the Internet be if routes were chosen to minimize latency?  Previous measurement campaigns suggest the following rule of thumb for estimating Internet latency: multiply line-of-sight distance by 2.1, then divide by the speed of light in fiber.  But a simple computation of shortest-path lengths through the conduits in the InterTubes map suggests that the conversion factor for all pairs of the 120 largest population centers in the U.S.\ could be reduced from 2.1 to 1.3, in the median, even using less than half of the links.  To determine whether an overlay network could be used to provide shortest paths, and how well it would perform, we used the diverse server deployment of a CDN to measure latency across individual conduits.  We were surprised to find, however, that latencies are sometimes much higher than would be predicted by conduit length alone. To understand why, we report findings from our analysis of network latency data from the backbones of two Tier-1 ISPs, two scientific and research networks, and the recently built fiber backbone of a CDN.}

\section{Introduction}



The Internet is built using hundreds of thousands of miles of optical fiber. This already extensive fiber outlay is also being continually augmented. In particular, several active and recently completed fiber projects are aimed at reducing Internet latencies along crucial routes~\cite{hiberniaExpress,zayo-seattle-chicago, arcticCable,chi-ny}.  While these infrastructural improvements are a welcome step, the Internet is known to make sub-optimal use of even its existing fiber~\cite{soslow-pam17}. A key contributor to latency inflation is Internet routing policy, which is driven by ISPs preferring cheaper transit routes, rather than minimizing latency~\cite{Spring:2003}.  In this context, a natural thought experiment comes to mind: how much could latencies be reduced if shortest paths were always followed?  Thanks to recent progress in mapping the Internet's fiber infrastructure~\cite{InterTubes}\footnote{We are indebted to the authors of the InterTubes work~\cite{InterTubes}, whose painstaking efforts in mapping US fiber infrastructure provided a starting point for our work. Their patience with our queries has also been commendable.}, we can examine this question in depth for at least the US geography, and glean insights that are more broadly applicable.

Our results show that utilizing the entire mapped US fiber plant in this manner would enable the design of a network interconnecting the 120 largest population centers in the US with (median) end-end fiber distances within $33\%$ of the geodesic distances. Recent measurement works have shown that latencies on the Internet are in the median $3.1 \times$ slower than the latency lower bound imposed by the speed-of-light~\cite{soslow-pam17,singla14hotnets}. Since the speed-of-light inside fiber is roughly $2/3$ of its speed in vacuum, this translates into a simple rule of thumb to convert geodesic distances to latency: multiply the distance by $2.1$ and divide by the speed-of-light in fiber, which is consistent with observations made in previous work on geolocation, \eg see Fig.~4 in the paper describing Octant~\cite{Octant}. In short, building this network by utilizing the available fiber could reduce the conversion factor from $2.1$ to $1.33$, which would be a very significant reduction in latency. We provide a first-order cost estimate for building such a network by leasing fiber-optic links. 


Building and operating a network on this scale, however, is a major undertaking. The obvious alternative is to deploy an \emph{overlay}, whereby suitable via points are used to transit traffic, such that the route through the via points is shorter than the default route~\cite{RON-Andersen,AkaOverlay-Ramesh1,AkaOverlay-Ramesh2}. Such overlays come at a cost: instead of paying for transit only at the end points, the overlay must pay for transit at each via point, and for deploying infrastructure for redirecting the overlay traffic appropriately. This trade-off is often favorable, however, and is frequently made by content delivery networks (CDNs)~\cite{theAkamaiNetwork}.

 So, what if we deployed overlay nodes at every fiber conduit's end points? Such a deployment would allow latency-optimal use of all existing fiber infrastructure.  To investigate this possibility, we performed extensive measurements between servers near opposite endpoints of fiber conduits using the infrastructure of a large CDN.  Our experiments revealed something unexpected: the measured latencies between endpoints near opposite ends of a fiber conduit are often significantly larger than expected based on the  conduit's known length. Further, the observed discrepancies are consistent and could not be explained by typical culprits, such as queuing delays. Thus, we also identify the root causes of such latency anomalies in the Internet's fiber infrastructure.

To uncover the factors that inflate measured latencies across individual conduits, we conducted an analysis of smaller scale measurements from the backbones of two large ISPs (AT\&T and CenturyLink), two scientific and research networks (ESnet and Internet2), and a recently built fiber backbone of a major CDN.  The CDN measurements were made by routers directly connected to leased long-haul fiber, rather than servers. 
Our main findings are:

\vspace{-0.06in}
\begin{itemize}\setlength\itemsep{0em}
    \item The fiber footprint in the contiguous United States is large and diverse and could be used to connect the 120 largest population centers with median stretch within $33\%$ of geodesic distances using less than half of the fiber links in the InterTubes map.
    
    \item Merely having servers near the endpoints of a conduit is not enough to guarantee low round-trip times (RTTs).  Despite a major CDN having servers near both endpoints of $51\%$ of conduits in the InterTubes map, we observed RTTs within $25\%$ of ``$f$-latency'' (the latency based on traversing the conduit length at the speed of light in fiber) only for $11\%$ of the conduits. Measurements we conducted on CenturyLink, and measurements published by AT\&T also support this finding.

    \item In some instances, fiber providers deliberately {\em increase} the amount of fiber in a link (through the addition of fiber spools) for the purpose of service differentiation.
    
    \item Using the published fiber routes of a large ISP and fiber link lengths obtained from a CDN backbone, ESnet, and Internet2, we performed a partial validation of the link lengths published in the recent InterTubes dataset. Even though for a large number of links, the differences we observe are small, there are also many cases where the lengths we obtain differ significantly from the published lengths in the dataset. We have also observed multiple fiber routes between pairs of cities, which partially explains the observed differences.
\end{itemize}

\begin{figure*}[t]
	\centering
	\subfigure[]{\label{fig:length-los-ratio}
		\includegraphics[width=2.2in]{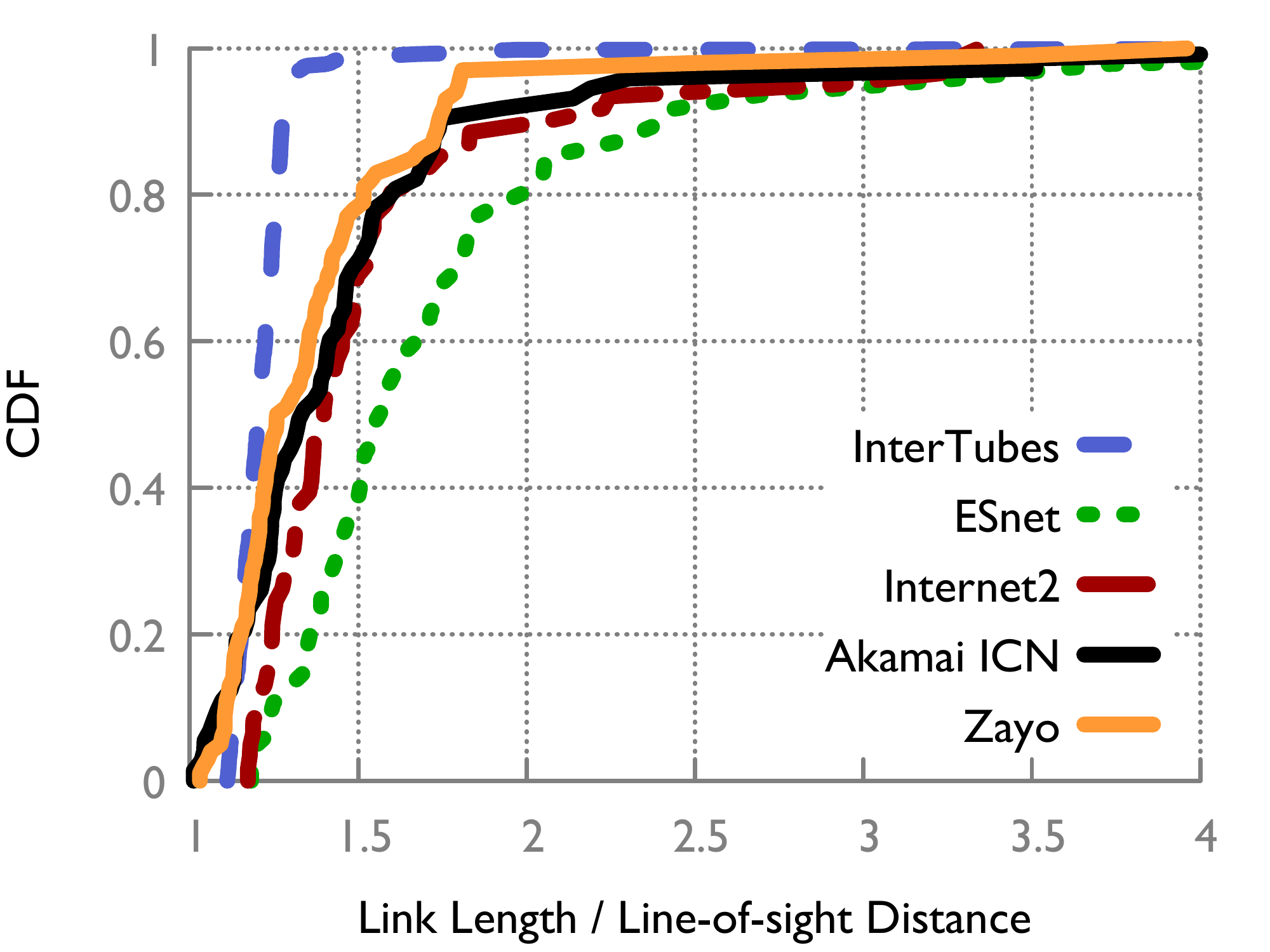}}
	\subfigure[]{\label{fig:length-los-scatter}
		\includegraphics[width=2.2in]{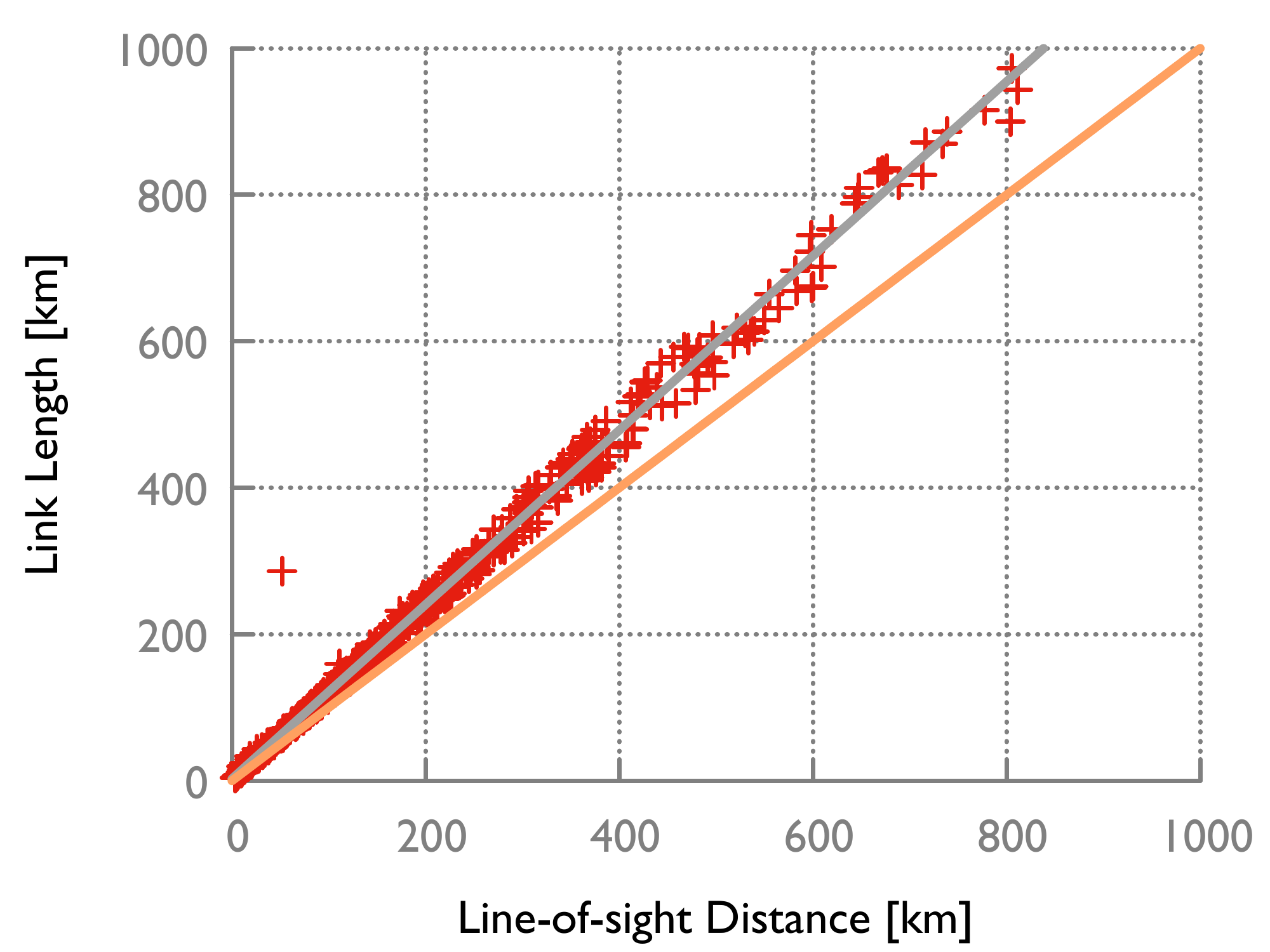}}
	\subfigure[]{\label{fig:unverified-links}
		\includegraphics[width=2.2in]{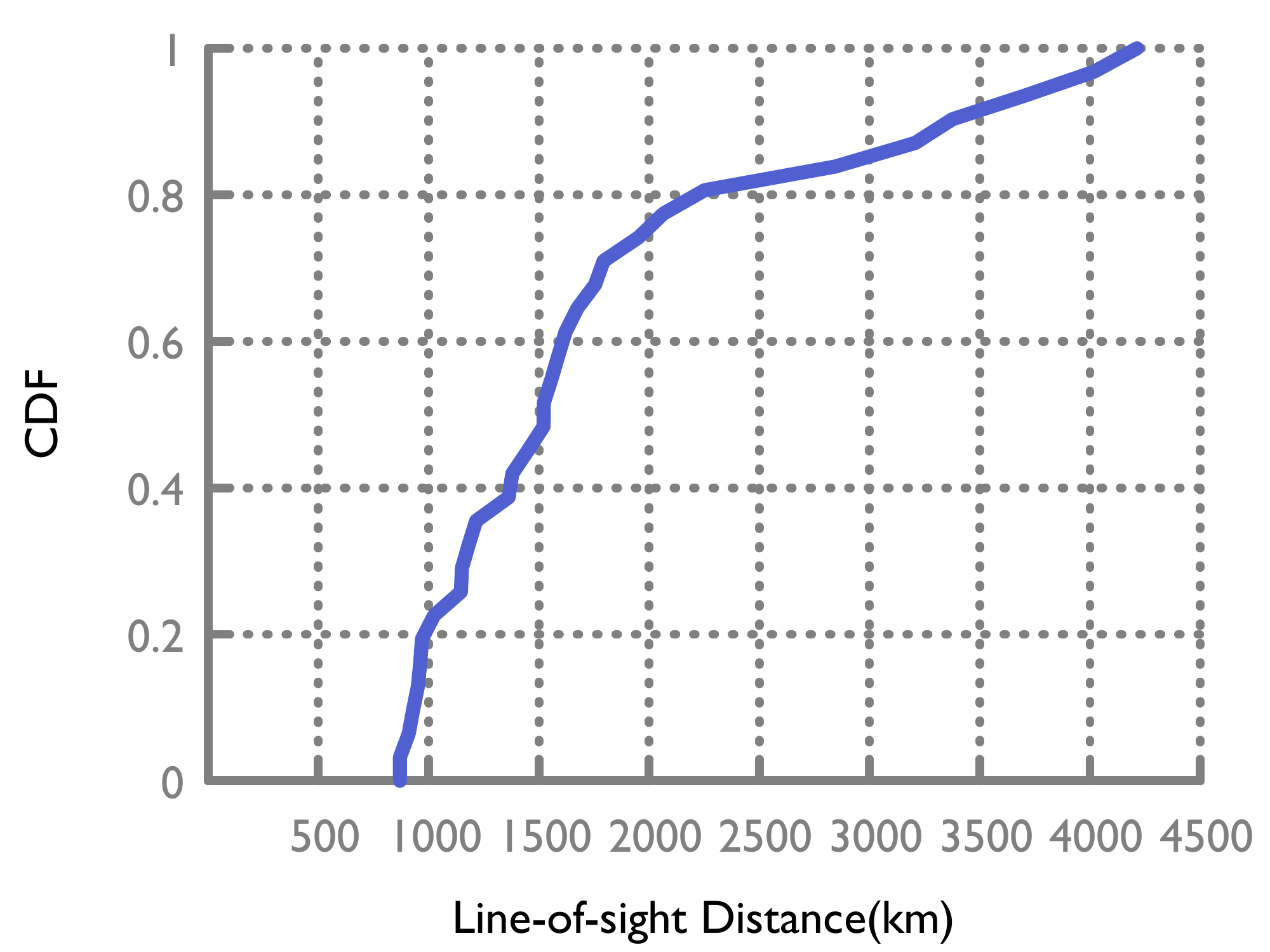}}
	\caption{Comparison of conduit lengths to line-of-sight distances}
	\label{fig:fiber-lengths}
\end{figure*}

\section{The InterTubes Dataset}
\label{sec:intertubes}

Durairajan et al.~\cite{InterTubes} compiled various public sources of information such as
network maps of large ISPs, government records, and contracts related to rights of way to analyze US fiber 
infrastructure. 
The resulting fiber map (henceforth, the InterTubes dataset) contains $273$ endpoints connected by $540$ long-haul links owned or used by 20 ISPs. 



We obtained conduit lengths for the InterTubes map from its authors. The median conduit length is $167$~km and roughly two-thirds of the conduits are longer than $100$~km. For $31$ links ($\sim$$6\%$), the conduit paths are unknown, so line-of-sight (LOS) distances\footnote{More precisely, geodesic distances: shortest paths between the endpoints along the Earth's surface.} were provided to us as a first-order approximation.  
Excluding these links, the dashed blue curve in Fig~\ref{fig:length-los-ratio} compares the conduit lengths to LOS. Conduit lengths are $20\%$ ($28\%$) longer than LOS in the median ($95$-th percentile). Results for links of individual ISPs are similar.


Figure~\ref{fig:length-los-scatter} compares the conduit lengths and LOS between individual conduit endpoints. We see that the gap between the points and the diagonal (\ie $y = x$) increases with link length, and for a small number of links ($8$), the difference exceeds $150$~km. The best-fit line has slope $1.19$, which is very close to the median ratio of conduit length to LOS ($1.2$).
Fig~\ref{fig:length-los-scatter} has no conduits longer than $~1,000$ km. This stems from the $6\%$ excluded links: Fig~\ref{fig:unverified-links} shows the LOS distances between the endpoints of these links, which are all long, and mostly run East-West. 
While routes for these conduits likely include other segments already in the dataset, this does not help estimate their end-end lengths because there are many possibilities of connectivity through these segments. This reveals an unfortunate limitation of the data: while the fraction of conduits of unknown length ($6\%$) is small, the fraction of \emph{distance} covered by these conduits is not: the lengths of these $6\%$ of links comprise fully $33\%$ of the total fiber miles. This calculation is an underestimate, computed using the minimal possible lengths for these $6\%$ of links (\ie LOS distances). 

Given the close agreement of most links around the median length-to-LOS ratio of $1.2\times$, the rest of our analysis estimates the lengths of these $6\%$ links as $1.2\times$ LOS.



\section{Making the most of fiber}
\label{sec:fiberDesign}

The InterTubes data raises an interesting question: how much faster would the Internet be if we could use \emph{all} available fiber unencumbered by Internet routing? Recent work~\cite{soslow-pam17} estimated that Internet path inflation compared to LOS distances is $2.1\times$ in the median\footnote{That work uses speed of light in vacuum as the baseline, resulting in a $3.1\times$ estimate; $2.1\times$ is simply that estimate adjusted to the speed of light in fiber.}. Given the value of reducing latency~\cite{soslow-pam17}, it is thus interesting to ask: how much of this $2.1\times$ inflation could be eliminated \emph{if} routing were not a constraint.

\subsection{Use all the fiber!}
\label{subsec:stretch}

For all pairs of endpoints in the InterTubes map, we compute the shortest fiber paths, and calculate their stretch compared to LOS. We find that the median ($95$-th percentile) stretch across all endpoint pairs is $1.32$ ($1.86$). Implicit in this calculation is the assumption that each pair of endpoints exchanges the same amount of traffic. Moving to a gravity model (\ie traffic between any two endpoints is proportional to their population product\footnote{We use population estimates from the 2013 census data.}), reduces the median ($95$-th percentile) stretch to $1.26$ ($1.56$).

The same analysis can be repeated for links available to each ISP separately. EarthLink and Level3\footnote{Now part of CenturyLink.} have the largest fiber coverage, with each associated with $\sim$$85\%$ of all conduits. As expected, the stretch for these two networks is very similar to that for the entire fiber map. ISPs with fewer fiber conduits, however, incur much larger stretch. For $8$ ISPs (out of $20$), the network is disconnected when restricted only to links identified as used by that ISP in the InterTubes dataset. This could result from conduits being missing or not attributed to ISPs in the dataset. For instance, advertised latencies between some locations in AT\&T's network~\cite{att-latency} are not achievable over their link set in InterTubes.

Gaps in the data notwithstanding, it is clear from past work's measured stretch ($2.1$ in the median)~\cite{soslow-pam17}, and the simple computation above ($1.32$ and $1.26$ in the median for the two traffic models) that being able to use all the available fiber would yield a large improvement in latencies of around $40\%$. More unmapped fiber could only make this possibility more enticing. Attempting to use the entirety of the fiber infrastructure, however, is likely to be cost sub-optimal, particularly because many of the endpoints in the InterTubes dataset are in sparsely populated areas.

\subsection{Supercharging inter-city connectivity}
\label{subsec:stretchLessfiber}

What if the big population centers could be all interconnected with the fastest or near-faster possible fiber routes? We zero in on $120$ large population centers in the US, obtained by coalescing smaller cities and suburbs near the $200$ most populous cities.
The mean distance between these population centers and their corresponding nearest fiber conduit endpoints is small ($5.1$~km), so we use these nearby fiber endpoints and our population centers interchangeably. We find that a fiber network using all the edges on the shortest paths between all pairs of these $120$ target endpoints uses $348$ links, with $137,500$~km of fiber, and achieves a median stretch of $1.28$ (in the gravity model, which we use throughout henceforth).

We investigate the impact of using fewer links using a simple iterative heuristic: at each step, all links that are not cuts are tested, and the link which would, if removed, cause the minimum increase in the mean stretch, is deleted.

\begin{figure}[t]
	\centering
	\includegraphics[width=3.2in]{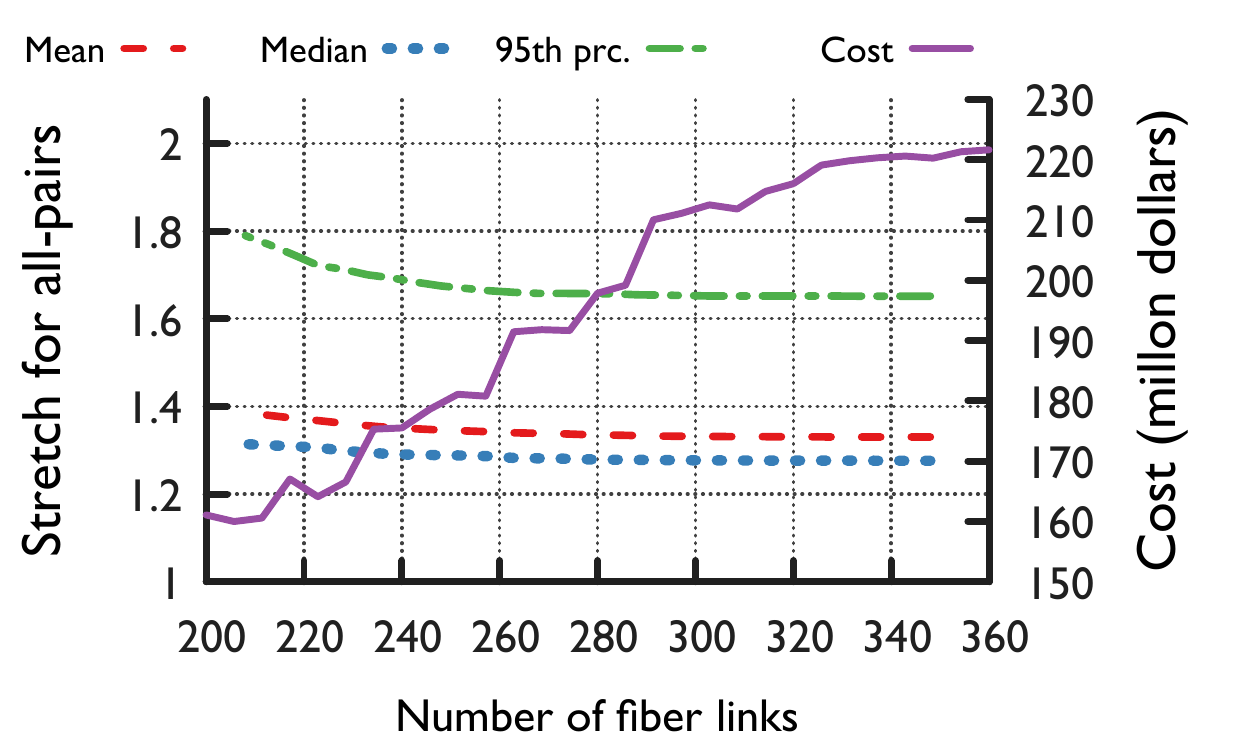}
	\caption{The trade-off between the number of links in the fiber topology, and stretch and cost of the 
		network}
	\label{fig:fiber-cost-stretch}
\end{figure}

\begin{figure*}
	\centering
	\includegraphics[width=6.4in]{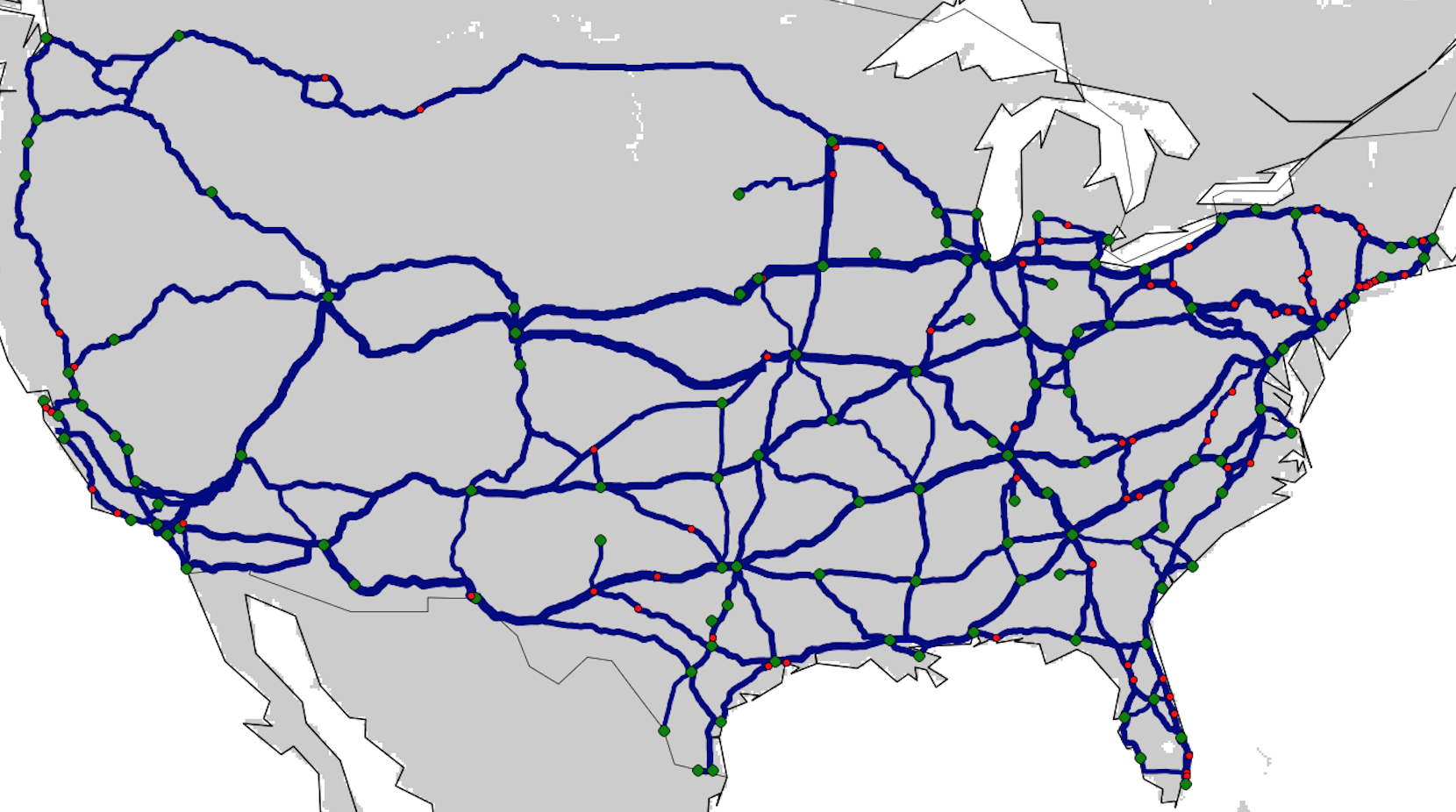}
	\caption{The fiber network which connects the $120$ population centers with $268$ links and $82,000$km of fiber.
		Increasing link capacities along the heavy traffic routes are marked with darker colors and thicker lines. Green circles represent the $120$ population centers and red circles represent conduit endpoints.}
	\label{fig:fiber-design}
\end{figure*}

Fig.~\ref{fig:fiber-cost-stretch} shows the mean, median, and $95$-th percentile stretch for networks built with different numbers of fiber links.
Stretch does not change appreciably until a large number of links are deleted (leftward on the $x$-axis). 
One example network with $268$ links using $82,000$ km of fiber is shown in Fig.~\ref{fig:fiber-design}.
The median, mean and $95$-th percentile stretch values for this 
network are $1.28$, $1.34$ and $1.66$ respectively.
Thicker lines show higher utilization under shortest path routing with the gravity model for traffic.
The map only draws actual fiber routes where available ($\sim$$40\%$ of links); the rest are represented instead by roads between link endpoints. 

\parab{Cost estimation:} What would this fiber map cost to be able to use?
We use a $1$~Tbps aggregate input rate to address this question, but arrive at a per unit bandwidth estimate that does not change substantially with aggregate input rate.
Our cost analysis is based on leasing the necessary fiber links, specifically, using short term\footnote{$3$-$5$ years in this context.} wavelength leases.
Typically, wavelength services are available at $1$, $10$, $40$ and, $100$ Gbps, and prices vary 
depending on bandwidth, distance, route, carrier, and contract length. For a few key routes such as New York to Los Angeles, and, New York to
Chicago, we found pricing for $10$ and $100$ Gbps wavelengths in the form of $\$ / (\text{Gbps} * \text{km})/\text{month}$~\cite{100G-price,pricingTrends}. For 
example, the median $100$~G wavelength price for the New York - Chicago is $\sim$$\$30,000$ per month (anecdotally, much lower prices can be negotiated), and the distance 
is ~$1,200$ km, computing to $0.25$ $\$ / (\text{Gbps} * \text{km})/\text{month}$. This price 
estimate is for a $5$ year contract. For each site, we estimate an equipment cost of $\$10,000$ and a monthly 
co-location fee of $\$2,000$ for hosting this equipment.


%


Using an aggregate input rate of $1\text{Tbps}$, we computed the bandwidth needed per link for shortest-path routing.
For each link, we picked the wavelength capacity such that utilization would be $20$-$90\%$. For some links, this
meant using $2$ wavelengths of the same capacity.
Figure~\ref{fig:fiber-cost-stretch} shows
variation in cost with the number of network edges. Overall, cost is smaller with fewer edges, but as the network gets 
smaller, the increasing load on 
some links requires upgrading to higher capacity wavelengths, and as an artifact of this, we see 
higher costs for some networks with fewer edges. 

Based on a $5$-year lease, the cost of the network in figure~\ref{fig:fiber-design} is estimated to be 
$\$190$ million. This cost is dominated by bandwidth costs, which total to $\$168$ million over $5$ years; the rest being equipment and co-location cost.  
Amortized over the network's goodput over $5$ years, this cost estimate translates to $\$ 0.009 / \text{GB}$.
Even though this seems very low at first glance, it is about $10\times$ more expensive than the New York to 
Chicago $100~\text{Gbps}$ link. This increase comes from substantial unused capacity on many links, and the average hop count of $9$ links on end-end paths. 

Our objective is to establish a first-order cost estimate, so we omit an analysis of different traffic models here. Beyond cost, the most interesting question is: how do we translate these calculations and estimates into practice?


\section{Bootstrapping off a CDN}
\label{sec:cdn}


The results of the previous section were based on fiber availability and length estimates in the InterTubes data set. Are these results achievable in practice? 
Deploying and operating equipment at a large number of fiber conduit endpoints is a massive undertaking. Conveniently, content distribution networks already operate points of presence across many fiber routes, and we are fortunate to be able to make use of a major CDN's server infrastructure and network connectivity to collect measurements with the goal of empirically upper-bounding the achievable latency on observable routes. 

\subsection{Measurement setup}
\label{sec:cdn-overlay-setup}

We identified clusters of the CDN near (more precisely, within $25$~km) the fiber conduit endpoints in the InterTubes dataset. We were able to find clusters near $51\%$ of the conduit endpoints. We picked at least one cluster in each of these locations, and used one server in each cluster for measurements. If the CDN uses multiple ISPs for network connectivity at a certain location, servers in more than one cluster are used to cover all ISPs for measurements. For each conduit for which we had servers near both endpoints, we ran traceroutes between all pairs of servers near the two endpoints of the conduit. Considering two directions separately, we ran traceroutes across $552$ links (out of $ 2 \times 540 = 1,080$).


Our objective with these measurements is to \textit{cover} as many fiber links between CDN servers as possible, and try to observe latencies within a small percentage of $f$-latency. This also necessitates that the internal latency within each endpoint site be small. Thus, we also ran traceroutes between all pairs of servers at each location.


Considering two directions along each conduit separately, and including measurements between servers  near the same conduit endpoint, we ran traceroutes between $82,382$ server pairs.
Traceroutes were run every $\sim$$3$ hours for $2$ days on two different occasions. Only $3.6\%$ of these measurements were found to be between servers which have the same network provider, covering $125$ fiber links. The rest of the measurements are between servers with different network providers, along $543$ fiber links. Note that we are counting each fiber conduit separately for each direction of traffic.

\subsection{Summary of the results}
\label{sec:cdn-results-summary}
We recorded the minimum measured RTT observed in the traceroutes for each pair of servers, and compared it to the $f$-latency between their locations. We used $c = \ukmps{204000}$ as the speed of light in fiber.

Even if a traceroute traverses an intended conduit, additional latency beyond that expected over this conduit may be incurred because of the (small) distance of the clusters from the conduit endpoints. (Recall that the clusters are chosen such that they are within $25$~km of conduit endpoints.) If paths between servers and their nearby conduit endpoints follow LOS fiber links, and the endpoint-to-endpoint path is the intended fiber conduit, the server-conduit distances can increase the RTT beyond the $f$-latency by at most $0.5$~ms.

Observed mininum RTTs, however, exhibit much larger deviations from their $f$-latency targets.
Only $124$ fiber links have RTTs within $25\%$ or $0.5$~ms (whichever is larger) of $f$-latency, corresponding to $11\%$ of all links, and, only $182$ fiber links have RTTs within $40\%$ or $0.5$ ms of $f$-latency, corresponding to $17\%$ of links. 

We also found that RTTs between clusters around the same location were higher than expected. 
For locations with many clusters, the median inter-cluster latency is small (a few ms), but for many locations which have a small number of clusters, the median inter-cluster latency is often in the $10$-$30$~ms range. 
\subsection{On unexpected measurement results}
\label{sec:cdn-latency-discussion}


A key difficulty in identifying causes of deviations from the expected $f$-latency for many cluster pairs is the lack of our visibility below layer $3$.

Traceroutes help us to identify cases where the inflated RTT is due to long, circuitous paths caused by sub-optimal routes which might happen due to a number of well known reasons~\cite{Spring:2003}.
However, for many conduits, we have a large number of pair-wise measurements along the conduit utilizing all the different network providers on both ends at distinct cluster locations. Even when the two servers were in the same network, which we hoped would lead to direct, shortest paths to be taken, we observed that measured RTTs were much higher compared to $f$-latency. For those measurements, the paths indicated by traceroute do not indicate any detours in many cases; yet, we still cannot be sure whether the packets followed the intended fiber conduit. The use of MPLS traffic engineering or the lack of direct optical circuits may cause longer physical paths to be taken.

To make the discussion above more concrete and show the difficulties we had with explaining the high RTTs, we focus on the server pairs with the minimum RTT for each fiber conduit. For each such server pair, we geolocated the routers in the traceroute output. To identify measurements which potentially took the intended fiber conduit we used the following heuristic: After geolocating the routers, we identified the consecutive pair of routers $(r_1,r_2)$ with the maximum distance among all consecutive routers. Traceroutes in which $r_1$ and $r_2$ are within a small distance of the conduit endpoints and the geodesic distance between $r_1$ and $r_2$, $d(r_1,r_2)$, is close to the fiber conduit length are good candidates for measurements in which the path uses the intended fiber conduit. 

Using the RTTs, we filtered the cases with obvious geolocation errors, and identified $r_1$ and $r_2$ for the remaining traceroutes. Using a $15$ km threshold for $r_1$ and $r_2$'s distance to the endpoints and picking only the traceroutes in which $d(r_1,r_2)$ is within $30\%$ of the conduit length, we ended up with $84$ traceroutes. Note that each of these $84$ traceroutes has the minimum RTT among all measurements for that server pair and that particular server pair has the lowest minimum RTT among all server pairs along the corresponding fiber conduit. Even for these $84$ traceroutes, measured RTT is $59\%$ over $f$-latency in the median. 

One of the many similar examples with RTT significantly higher than $f$-latency is for the fiber conduit between Salt Lake City (SLC) and Phoenix (PHX). The conduit length is $943.6$ km, computing to a $9.25$ ms $f$-latency.
For both directions, the same server pair has the minimum RTT ($15.1$ ms) among all pairs, which is $63\%$ over $f$-latency. We also observed that the routes are symmetric, so we just describe it for one direction.
From PHX to SLC, we identify $r_1$ and $r_2$ as the $4$-th and $5$-th hops respectively.
Both $r_1$ and $r_2$ are within $1$ km of the conduit endpoints and the RTT to $r_1$ is $0.4$ ms. 
Minimum RTT to $r_2$ is $15$ ms among all traceroutes, and we see that $d(r_1,r_2)$ is $811.2$ km. 
Conduit length is $16\%$ longer than LOS distance between the endpoints (and $r_1$ and $r_2$), which is consistent with our earlier analysis of the conduit lengths, yet minimum RTT has $63\%$ inflation over the $f$-latency computed from the conduit length. 

The example above is for the best server pair -- i.e., the one with minimum RTT -- among many, between SLC and PHX, and the layer 3 path indicated by traceroute can be said to be optimal. Examining the measurements between other server pairs along the same fiber conduit helps illuminate why we were able to cover only a very small percentage of fiber conduits with the overlay: despite having measurements for tens of thousands of server pairs, most measurements use suboptimal paths. For this SLC-PHX fiber conduit, we have $50$ unique server (cluster) pairs for measurements, $25$ in each direction. Using the minimum RTT for each server pair, the median RTT across server pairs is $4.6 \times$ the $f$-latency. Even the second best server pair's minimum RTT is $2.8\times$ the $f$-latency and passes through routers in Los Angeles and San Jose, which clearly indicates the direct fiber conduit is not taken. For other server pairs, we observe longer detours in the router paths. So, for this conduit we had $1$ server pair out of $25$ in each direction which potentially used the fiber conduit, and even for that pair we observe $63\%$ inflation over $f$-latency.
\section{What went wrong?}
\label{sec:fiber-latency}

The measurements between the CDN servers showed us that even when there are no indications that the direct fiber link is not taken, observed RTTs show significant inflation over $f$-latency. Even though one might speculate that longer MPLS tunnels might be the reason of observing higher RTTs in some cases\footnote{Not that we have any proof of or any way of investigating that with our collected data.}, it is reasonable to look into other potential factors since we had measurements between a large number of pairs of servers with traceroutes. It might very well be that we were optimistic about the contribution of factors related to lower layers and their contribution should have been taken into account. Another explanation might be that in some cases the conduit lengths, or the cable lengths inside conduits, are actually longer than reported in the InterTubes dataset that we relied on, and as a result, we under-estimated $f$-latency. Before delving into a more detailed look into these factors with both fiber length information and latency measurements from other networks, we examine the most significant contributors of latency to an optical signal between a pair of locations.

\subsection{Sources of latency inside fiber}
\label{sec:fiber-latency-detailed}

\paragraphb{Speed of light inside fiber:} The speed of light inside fiber is roughly $2/3$ of its speed in vacuum, and we have used $c = \ukmps{204000}$ for speed-of-light inside fiber to be exact for all analysis in this paper. Even though this choice is extremely unlikely to be a factor affecting our analysis and does not make a practical difference for many applications and distances, it is worth mentioning that different fiber-optic cable designs and specifications have slightly different refraction indices and hence propagation speeds.
Refraction index quantifies the slowdown of the propagation speed of light inside fiber compared to vacuum.
For example, refraction indices for an ``Ultra-Low Latency Fiber'', standard Single Mode Fiber (ITU G.652 SMF), and commercial Non-zero dispersion-shifted fibers (ITU G.655 NZ-DSF) are given as $1.462$, $1.468$, and $1.470$ respectively by a commercial fiber provider~\cite{corning}, resulting in propagation speeds $205.1$, $204.2$, and $203.9$ km/ms. 
Even though the differences in propagation speeds are small, for applications like high-frequency trading the latency differences are significant. Spread Networks, for instance, has a $1328$ km fiber link between New York and Chicago; for this distance the ultra-low latency fiber leads to $71$ $\mu$s reduction in RTT compared to a typical commercial NZ-DSF fiber. 

Interestingly, Spread Networks mentions that they use TrueWave RS Optical Fiber, a state-of-the-art NZ-DSF fiber conforming to ITU G.655.C, and its (group) refraction index is given as $1.47$ in the product description~\cite{true-wave-rs}. Hence, it has the slowest propagation speed among the listed fiber types above. We briefly discuss at the end of this section what the advantages of NZ-DSF fibers are, leading to this choice. Even though relatively recently laid fiber for new applications such as high-frequency-trading might use fiber-optic cables utilizing latest transmission technology, most of the fiber underground is old, standard SMF according to Steenbergen~\cite{opticalTutorial}. The work of Filer et al.\ about elastic optical networking in Microsoft's backbone includes a breakdown of the fiber-optic cable types in their network~\cite{elasticOptical} and, interestingly, only 7$\%$ is classic SMF, whereas almost 90$\%$ consist of variations of G.655 NZ-DSF fibers.


\paragraphb{Fiber path (conduit length):} Reducing the physical path length the fiber follows as much as possible is of utmost importance, since propagation delay is the largest contributor to signal latency between two locations. It is well known that Spread Networks reduced the distance between New York and Chicago stock exchanges by laying out new fiber along shorter and shorter routes successively, culminating in ``the straightest and shortest route possible'' as advertised on their website~\cite{spread-networks}. In general, we know that fiber paths almost never follow a straight line between two locations as we quantified in earlier work~\cite{soslow-pam17,singla14hotnets} for a few networks, and as analyzed for conduit lengths in US mainland provided in the InterTubes dataset in \S\ref{sec:intertubes}. 
We shall later compare the lengths of common links in the InterTubes dataset and a few other networks for which we obtained fiber lengths. Moreover, we obtained the detailed long-haul and metro area fiber routes from one ISP, and we used this data to cross-check the lengths in the InterTubes dataset.

\paragraphb{Slack loops and tube design:} It is important to note that the length of the fiber cable is longer than the length of the conduit (tube) it is contained, and the difference in lengths is non-negligible. When new fiber is laid out in the ground, some excess fiber needs to be set aside for fiber cuts and required repairs. It is not possible to pull fiber over long distances, and a small amount of extra fiber is, hence, placed in slack loops at regular short intervals, \eg every 200 m.  
A construction manual from a cable provider and a tutorial about fiber installation recommend 
leaving slack loops at least totaling 5$\%$ of the cable length~\cite{fiberSlack2,fiberSlack1}. 
The previously mentioned Chicago - New York Spread Networks link is $\ukm{1328}$ long including the slack coils~\cite{spread-networks}. We computed the length of this link as $1253$~km from its physical path on the ground, and this implies that the amount of slack fiber left during construction is roughly $6\%$.

Moreover, the tube and cable design can have a noticeable impact on excess fiber length as well, and can increase the signal latency. For example, depending on the choice of a loose tube cable design or central tube ribbon cable design, and the number of fiber strands in the tube, excess fiber length can be anywhere between $0.3\%$ and $7.9\%$ of the total cable length as calculated in~\cite{low-signal-latency}. 

\paragraphb{Opto-electrical components:} Long haul fiber links provide connectivity between backbone routers inside ISP PoPs. In the common IP-over-WDM (Wavelength Division Multiplexing) architecture, transponders/muxponders connected to the backbone routers will convert electrical signals to optical, to be transmitted out of ROADM (reconfigurable optical add-drop multiplexer) or OXC (optical-cross-connect) devices. Depending on their complexity and capabilities, transponders/muxponders can cause an extra latency from a few $\text{ns}$ to $10-100$ $\mu s$~\cite{infinera,optelian}. Use of forward error correction (FEC) can result in an additional $15-150$ $\mu s$ latency~\cite{optelian}.

\paragraphb{Optical amplifiers:} The fiber connecting two ROADMs in two distant PoPs will go through a number of optical amplifiers to extend the reach of the optical signal. Most common types are  Erbium Doped Fiber Amplifiers (EDFA) and Raman amplifiers. EDFA uses a small amount of extra fiber for amplification in each unit, and passing through a large number of amplifiers in a long fiber link might cause a few $\mu s$ additional latency, which might be important for some cases. More importantly, use of many EDFAs in a WDM system will cause the power levels of different signals in the same fiber to diverge, which might necessitate optical signal regeneration, which will add to the latency. In general, as the number of network elements such as amplifiers, ROADMs or any other transport equipment that the signal needs to traverse increases, resulting latency overhead will increase.

\paragraphb{Chromatic dispersion:} The most significant source of latency in the optical components of a signal's path is chromatic dispersion. Different wavelengths 
have very slight differences in speed, causing the signals to spread out as the traversed distance increases. 
Dispersion compensation fibers (DCF) are utilized to negate this effect, very commonly using extra fiber spools with negative dispersion at optical amplifiers. Hence, using DCF increases the total fiber length, resulting in $15-25\%$ latency overhead~\cite{bobrovs-dcf}. The amount of extra fiber used for DCF differs depending on the transmission technology and type of the fiber. While standard SMF has a higher ($20-25\%$) latency overhead due to DCF, ITU G.655 NZ-DSF fibers have better dispersion characteristics and typically add around $5\%$ to the fiber length and latency. 

In practice, exact DCF overhead can be computed based on the cable specification and dispersion compensation coefficient, with units in $\text{ps} / \text{nm} * \text{km}$. The dispersion compensation coefficients of standard SMF and a commercial NZ-DSF fiber is given as $16.5$ and $4.2$ $\text{ps} / \text{nm} * \text{km}$ respectively~\cite{dcf-survey}, implying a $75\%$ reduction in DCF overhead with modern fibers compared to standard SMF. Moreover, dispersion compensation can also be performed using Fiber Bragg Grating technology, which does not require extra fiber spools, and effectively removes the latency overhead due to dispersion compensation~\cite{infinera}.

\paragraphb{Publication of ``mock'' routes:} In personal communications, a fiber provider informed us that published maps contain routes simply produced from Google Maps driving directions.  The rationale for publishing these ``mock'' routes is to avoid revealing competitive details. We are unaware, however, of any instance in which the resulting length differs greatly from the true conduit length.

\paragraphb{Fiber added to increase latency:}  To our surprise, we discovered that sometimes additional slack is purposefully added for price differentiation, such as Hibernia adding $440$ km slack to the New York - London route to reduce prices for non-HFT customers~\cite{stronge-slack}. In personal communications with a fiber provider, we learned that such slack has been added to at least one fiber route in the US mainland as well.

\subsection{Summary: sources of optical signal latency}
\label{latency-sources-summary}
Latency over a fiber-optic link can accumulate due to many reasons. It is essential to know the network elements and the type and characteristics of the fiber-optic cable to have an accurate estimate of the expected latency. However, the most significant factors are length of the fiber path on the ground, amount of excess fiber due to slack loops and cable design, and additional fiber due to dispersion compensation, usually in amplifiers. Revisiting the latency optimized Chicago - New York fiber link, the choice of the particular fiber type mentioned before despite not having the highest propagation speed compared to other mentioned commercial fibers is most likely due to its better dispersion and attenuation characteristics; not only reducing the latency overhead due to DCF, but also increasing the reach of the optical signal without amplification to 120 km~\cite{spread-networks}, minimizing the number of network elements on the signal's path.

In section~\ref{sec:cdn-latency-discussion} we discussed the measurements along the fiber conduit 
between Salt Lake City and Phoenix, with the observed $63\%$ inflation in minimum RTT over $f$-latency even for the best server pair, with a close to ideal path observed in the traceroute output. 
Considering the impact of excess fiber due to slack loops and cable design, we can maybe roughly account for \textasciitilde$10\%$ inflation. In addition to that, even if we assume $25\%$ overhead due to DCF in the worst case, we still cannot explain the observed $63\%$ inflation over $f$-latency.


So, for these two large ISPs we only examine the impact of conduit length, since we don't have visibility into the other factors. However, through personal communications with their engineers, we obtained information about the amount of DCF overhead in AT\&T's network. 




\section{Latency in ISP \& CDN Backbones}
\label{sec:isp-cdn-backbone}
We next examine RTTs between the major PoPs in AT\&T and CenturyLink, and Akamai's recently built backbone ICN. For AT\&T and CenturyLink, we used the fiber link lengths in the InterTubes dataset as the baseline for calculating $f$-latency, whereas for ICN, we have detailed fiber routes and hence ground truth conduit lengths.

\subsection{Latency in the AT\&T backbone}
\label{sec:att}

\begin{table*}[t]
	\centering
	\begin{tabular}{rlrrrr}
	\toprule
	\multirow{2}{*}{\bf City 1} & \multirow{2}{*}{\bf City 2} & {\bf Fiber Length} &  {\bf $f$-latency} & {\bf AT\&T min. RTT} & {\bf CDN min. RTT} \\
	& & (km) & (ms) & (ms) & (ms) \\
	\midrule
        Atlanta & Dallas & 1418.38 &13.91&17&18.32 \\
		Atlanta & Nashville &405.31&3.97&8&9.26 \\
		Dallas & Houston &446.69&4.38&5&6.55 \\
		Dallas & New Orleans &826.97&8.11&12&13.4 \\
		Houston & San Antonio &379.8&3.72&5&5.28 \\
		Indianapolis & St. Louis &434.21&4.26&13&7.35 \\
		Kansas City & St. Louis &458.1&4.49&6&6.28 \\
		Nashville & St. Louis &460.99&4.52&12&12.85
	\end{tabular}
	\caption{Comparison of  latencies of the CDN and AT\&T over fiber conduits in AT\&T's backbone}
	\label{table:att-vs-cdn}
\end{table*}

AT\&T publishes latencies between major cities in its network on its Web site~\cite{att-latency}. 
We collected this data between February 27, 2017 and April 7, 2017 by recording it every 30 minutes. We obtained RTTs between $259$ city pairs for $24$ different cities. The methodology and details of AT\&T's measurements in their backbone is explained in~\cite{att-methodology,att-ieee}.

For each city pair with latency data, we computed the shortest path between them using  AT\&T's links in the InterTubes dataset. For $7$ city pairs, we observed that the RTT is lower than $f$-latency (based on the shortest path length), indicating that some links in AT\&T backbone are missing in the InterTubes dataset. This might also be due to some existing links in the dataset not marked as owned/used by AT\&T despite actually being in AT\&T's network. For example, the published RTT between Chicago and Indianapolis is $5$ ms, however the shortest path with AT\&T's links is $927.86$ km long, leading to a $9.1$ ms lower bound. There exists a $322.8$ km long fiber conduit between Chicago and Indianapolis (leading to $3.16$ ms $f$-latency), but this conduit is not marked as used by AT\&T. If we assume this link is in AT\&T's backbone, however, the resulting latency inflation over $f$-latency would be $58\%$.
For the remaining $252$ city pairs, AT\&T's latencies are $31\%$ over $f$-latency in the median. However, this might be misleading since the missing links will increase the shortest path lengths (and estimated $f$-latency), and result in lower inflation over $f$-latency. For comparison, published latencies are $82\%$ inflated over line-of-sight distances in the median. 

\begin{figure}[t]
	\centering
	\includegraphics[width=3.33in]{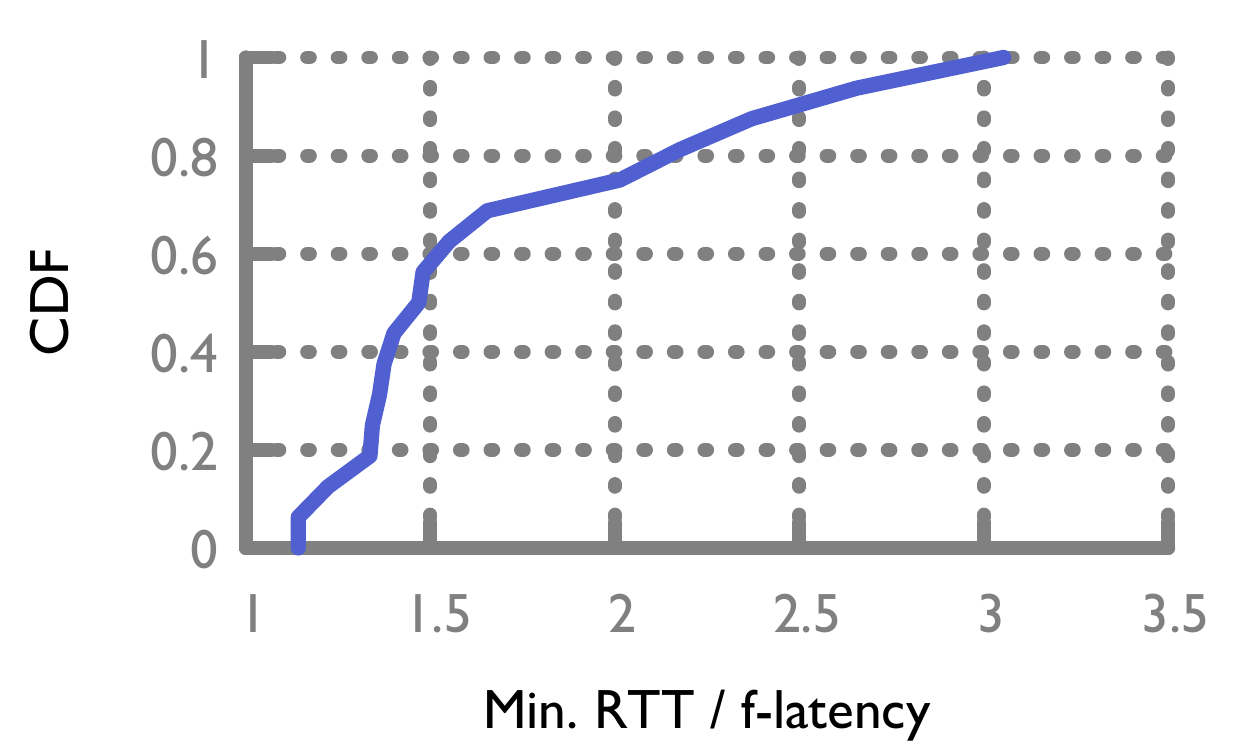} 
	\caption{CDF of latency inflation over 16 city pairs with direct fiber links in AT\&T network.}%
	\label{fig:att-latency}
\end{figure}

Due to the uncertainty about missing links and limited conclusions we can reach for the majority of published latencies, we next focus on city pairs for which 
\begin{enumerate*}[label=(\roman*)] 
\item we have RTT data and 
\item there is a direct fiber conduit between them in AT\&T's network according to InterTubes dataset.
\end{enumerate*}
We identified $16$ such city pairs. 
Figure~\ref{fig:att-latency} shows the CDF of the latency inflation over $f$-latency for these $16$ links. Minimum, median and maximum inflation are found as $1.14$, $1.47$ and $3.05$ respectively. Almost $1/3$ of the RTTs show more than $2 \times$ inflation! We also compared the latency across these links with the measurements with the CDN described in \S\ref{sec:cdn}. AT\&T is one of the network providers the CDN uses and some of our measurements were between CDN servers, both using AT\&T as the network provider. Table~\ref{table:att-vs-cdn} compares the RTTs in AT\&T and CDN data for $8$ city pairs with for which we have measurements in both. In general, latencies in the CDN data are slightly higher than observed in AT\&T data, except for Indianapolis and St. Louis pair. 

The slightly larger latencies in the CDN data can potentially be explained by two factors: 1- Our measurements between CDN servers were performed in $3-6$ hour intervals over a few days, with $~10$ measurements on average per city pair, whereas the AT\&T data covers roughly a 5-week period with data updated by AT\&T every 15 minutes; 2 - Each CDN server is selected using a $25$ km radius from the conduit endpoint locations, resulting in extra hops (and distance) before hitting AT\&T backbone routers. Regardless, latencies observed in both networks are substantially higher than $f$-latency, except for the Dallas-Houston pair where minimum latency in AT\&T data is only slightly higher ($14\%$) than $f$-latency. For example, both latencies between Atlanta and Nashville are more than $2\times$ inflated and the AT\&T's published latency between Indianapolis and St. Louis is almost $3\times$ inflated over $f$-latency.

We shared this data with AT\&T engineers to get their opinion about what might cause these large inflations.
We received a comment about some of the link lengths in our data being somewhat short. They also mentioned the impact of dispersion compensation as a potential factor; citing up to $17\%$ DCF overhead. This value is in the range of DCF overhead inside classic SMF as we described in section~\ref{sec:fiber-latency-detailed}, however it is not enough to explain the $2-3\times$ inflation observed for some links.

\subsection{Latency in CenturyLink}
\label{sec:centurylink}
CenturyLink  
provides a Web-based interface to run pings and traceroutes between some of its PoPs~\cite{centurylink-tool}.
We ran traceroutes, using this tool, between PoPs that are directly connected with a fiber conduit and for which we have the link length information in the InterTubes dataset.
We ran traceroutes between $16$ pairs of PoPs for $3.5$ months (between December 13, 2017 and March 27, 2018) once every hour. We observed path changes for only a few links and computed the minimum RTT for each observed path for all measured links. Using all $16$ pairs, we see that minimum RTT is inflated $51\%$ over $f$-latency in the median. 


There are $5$ links for which minimum RTT is more than $2\times$ the $f$-latency. One of them is over a short fiber link, and the absolute latency difference is less than $0.5$ ms. For the remaining four, we only see routers at the endpoint cities in all traceroutes.
Examining the IP/MPLS map of CenturyLink on its website, we observe that the endpoints of all these fiber links are in cities where MPLS nodes exist. Hence, longer MPLS tunnels not visible from traceroute output might be the cause of the observed RTTs. Also noteworthy is that in the optical wavelength map provided on the same page,\footnote{This optical wavelength map, which showed the established optical circuits between cities and their capacities, is no longer available at the same page.} for two of the links direct circuits are not available between the endpoints. For example, for the $664.7$ km long Boise - Portland link, we find $f$-latency to be $6.5$ ms whereas the minimum RTT is $13.5$ ms. The fiber map on CenturyLink Website shows this link, but the optical wavelength map does not. This map, however, shows circuits from both Portland and Boise to Tukwila,WA near Seattle, resulting in a path which is significantly (\textasciitilde$65\%$) longer, the use of which would explain the observed inflation in latency partially. 

To sum up, despite being able to find some plausible explanations for some of the large inflations observed thanks to the MPLS and optical wavelength information on CenturyLink's website, we still cannot account for large inflations for many of the links similar to AT\&T latency data.

\subsection{Latency in the Akamai ICN}
Akamai recently started operating its own backbone called Inter-City-Network (ICN)~\cite{icn-nanog}.
Inside the US, ICN connects $9$ major population centers using $19$ long-haul fiber links. We obtained the fiber routes underlying the ICN topology and computed the lengths of $74$ city-to-city fiber conduits. In a few cities, we have multiple fiber termination points. Before discussing measured latencies in ICN, we briefly examine the lengths of the fiber conduits in this network and compare them with the link lengths in the InterTubes dataset.

Using the locations of conduit endpoints, we computed the ratio of conduit length to LOS distance between conduit endpoints for each fiber conduit.The CDF of this ratio is shown in figure~\ref{fig:length-los-ratio} with the black curve in the middle.
The median of this ratio for ICN is $1.34$ compared to $1.2$ for the links in the InterTubes dataset. Moreover, for roughly $20\%$ of the conduits, this ratio is above $1.5$. This comparison of CDFs, however, is not over the same set of fiber conduits, and it might be misleading. When we compared the lengths of the common conduits between ICN and the InterTubes dataset, for half of the links with a common ISP the difference in lengths is within $10\%$. We also found a small number of conduits, however, for which the lengths in ICN are a few hundred km larger, up to $300$ km. 


Even though we have detailed information about the fiber-optic link routes, the information that we obtained is limited in the sense that we do not exactly know the physical path of each ICN link. Especially for the very long links which connect major cities in the West and East coasts, there are multiple plausible fiber paths with a few hundred km differences in length. Nevertheless, using latency tests we were able to identify the exact path for $10$ links.

We obtained measured RTTs across each ICN link in both directions. The difference in RTTs in the two directions is negligible for all links. For $6$ out of $10$ links for which we know the fiber routes exactly, measured RTTs are only $10-13 \%$ over $f$-latency. Moreover, for $5$ links the difference of lengths of alternative fiber paths are small compared to the length of each alternative route, and measured RTTs are within $8-17 \%$ of $f$-latency. For example, for the Dallas - Miami ICN link, the shortest fiber route is just over $3,300$ km and an alternative path is only $100$ km longer, corresponding to $11 \%$ and $8 \%$ inflation over $f$-latency respectively. For all these links, observed latency inflation can be explained by the excess fiber in slack loops and DCF overhead for NZ-DSF fibers.

There are a few links for which alternative fiber routes differ significantly in length, hence we are less certain about the inflation over $f$-latency. For example, measured RTT between Chicago and Atlanta is $37$ ms, and alternative fiber routes differ more than $600$ km in length, so resulting latency inflation over $f$-latency is between $18\%$ and $44\%$. However, for each such link there is at least one route which results in at most $24\%$ inflation over $f$-latency, which can be explained by the overhead due to slack loops and DCF in classic SMF if that particular route is underlying the ICN link. 


\subsection{Summary}
\label{summary-backbone-latency}
We find that latencies along the fiber conduits in both AT\&T and CenturyLink are higher than what we expected, with observed RTTs exceeding $2 \times f$-latency for some links. Our conversations with AT\&T engineers revealed DCF overhead as a contributor in addition to link lengths being under-estimated. In ICN, where we obtained the conduit lengths from the detailed fiber routes on the ground, we see that latencies are much closer to $f$-latency for a large number of links but we also observed significant differences in conduit lengths for a small number of links. To increase our confidence in the conduit lengths in the InterTubes dataset we next describe our work on verifying the lengths of fiber links in one large ISP's network, utilizing the detailed fiber routes published on its website.

\section{Verifying Zayo's Fiber backbone}
\label{sec:length-compare}

\begin{figure}[t]
	\centering
	\includegraphics[width=3.3in]{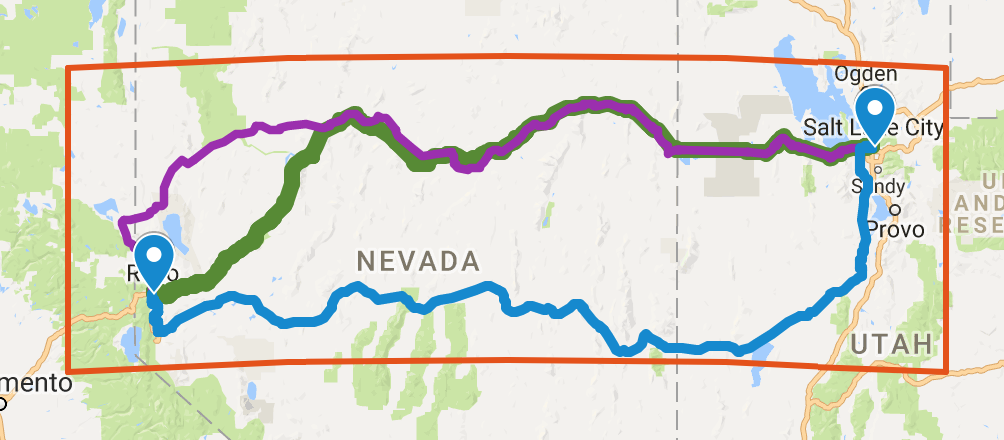}
	\caption{$3$ Reno-SLC paths over Zayo's backbone.}
	\label{fig:alternative-paths}%
\end{figure}

Zayo, a major ISP, publishes its fiber conduit routes on its website, in a KMZ\footnote{A compressed KML (Keyhole Markup Language) file.} file\cite{zayo-backbone}. This allows us to do a head-to-head comparison using Zayo's links in the InterTubes dataset. Using both the long-haul and metro area fiber routes and their coordinates in this file, we computed the lengths of the links in the InterTubes dataset that are listed as used/owned by Zayo. The file shows $112,854$ km of long-haul (in $28,745$ segments) and $57,776$ km of metro area (in $113,258$ segments) fiber in US mainland, either owned by Zayo or leased by Zayo from another ISP. 
%
To be able to stitch segments together and compute long-haul fiber lengths, we used the locations of the target endpoints in the InterTubes dataset to calculate bounding rectangles, and used only the segments which fall inside the computed boundaries to create a graph for finding paths between the endpoints in the InterTubes dataset. We omit the details of graph creation and path computation here, but we note that the segment endpoints almost never exactly match, so we had to assume nearby segment endpoints are identical if the distance between them is below a small threshold, often 10s of meters, though for a few cases we had to use values up to a few hundred meters.   


Using the generated graph we computed the length of the shortest path for each link in the InterTubes dataset and the lengths of a few alternative paths we observed in the visualized map. 
Figure~\ref{fig:alternative-paths} shows an example, where we computed the lengths of three distinct fiber routes between Reno and Salt Lake City. Length of this link is given as $813.7$ km in the InterTubes dataset, and we computed the lengths of three paths as $841$, $921$ and $1,019$ km. The figure also shows the locations of the two endpoints and the bounding rectangle used in filtering fiber segments and graph generation. 
%
We observe that for $20\%$ of the links, the length in the InterTubes dataset is larger than the length of the longest path we found. We also observe that for a small number cases even the shortest path length we computed is more than $1.5\times$ longer than the length in the dataset. In the median, the shortest, average and the longest path lengths are $8\%, 13.4\%,$ and $18\%$ longer than the length provided in the InterTubes dataset respectively. For \textasciitilde$12\%$ of the links, the length of the shortest and longest paths differ more than $100$ km.

\begin{figure}[t]
	\centering
	\includegraphics[width=3.3in]{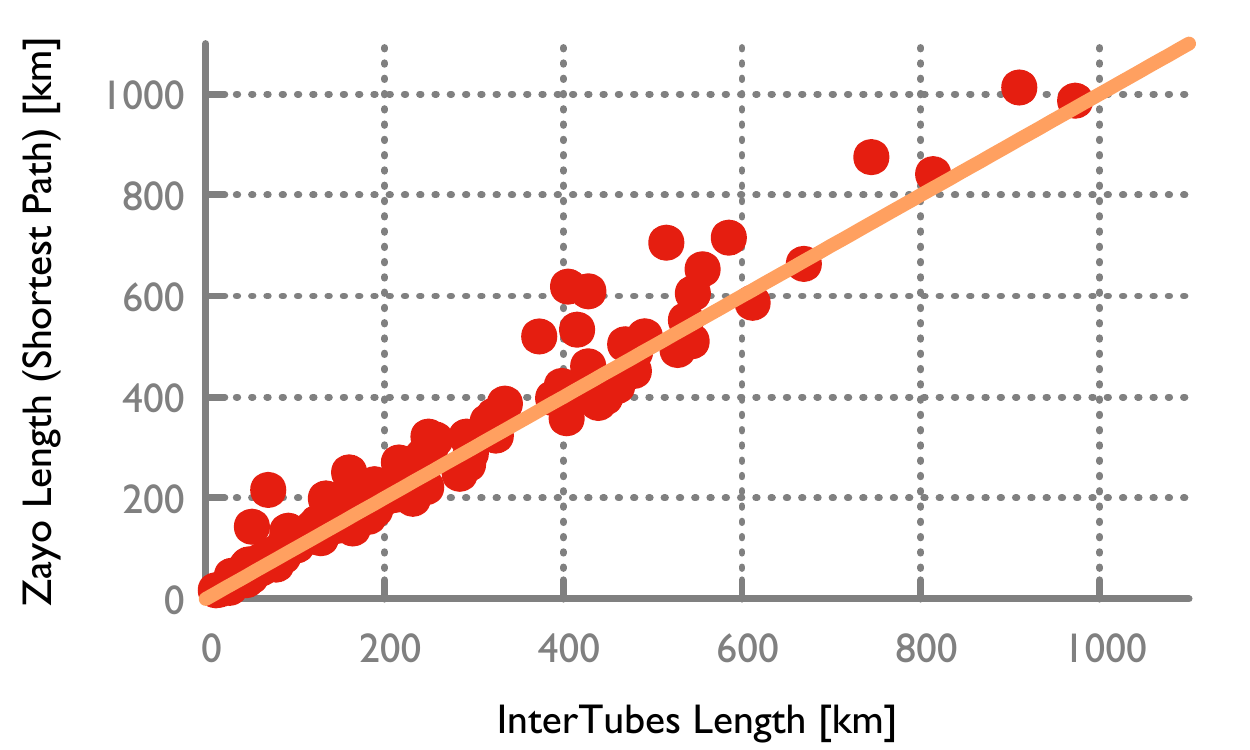} 
	\caption{Comparison of lengths in Zayo's backbone}%
	\label{fig:zayo-lengths}
\end{figure}

Figure~\ref{fig:zayo-lengths} shows a direct comparison of the lengths in the InterTubes dataset and the length of the shortest path we computed in Zayo's backbone. We observe a high concentration of points on or near the diagonal, especially for conduits up to $500$ km in length, though there is also a significant number of links where the difference in lengths is larger, where in all such cases the length we computed is longer. Since we have detailed coordinates of all the fiber segments as provided in the KMZ file by Zayo, larger observed differences might be due to the lack of such detailed knowledge of the routes for those links during the computation of the lengths in the InterTubes dataset. 

Overall, even though for a large number of links the difference in lengths we computed and what is in the dataset are small, there are many links for which the difference in lengths is significant which would cause us to undershoot while computing $f$-latency and result in a larger inflation. For example, in table~\ref{table:att-vs-cdn} we see that for the $405.3$ km Atlanta Nashville link, RTTs in the CDN and AT\&T data indicate more than $2 \times$ inflation over $f$-latency. InterTubes dataset shows both AT\&T and Zayo on this link, and from the Zayo file we computed its length as $618.7$ km, which is more than $50\%$ larger than the provided length in the dataset.

\section{Latency in ESnet and Internet2}
\label{sec:research-nets}

Internet2 and ESnet (Energy Sciences Network) share a large portion of their network 
since the last major upgrade of these networks, done in partnership with Level3 Communications \cite{esnet-internet2,level3-internet2}.
We obtained fiber link lengths and PoP locations from both networks. Some PoPs host just ROADMSs whereas others have also IP routers. From ESnet we obtained optical signal latencies between PoPs obtained with Optical Channel Laser Detectors and segment lengths estimated based on measured signal latency. Cross-checking this data by estimating latency across each link using $c = 204,000$ km/s in fiber, our estimates are on average within 45 $\mu$s of the measured latencies in the data, with the maximum difference being $66$ $\mu$s. From Internet2, we obtained the physical map of their backbone in an image file, which has markers showing PoP locations and distances between PoPs in kilometers. The distances in the map are provided to Internet2 by their network provider. 
We used the fact that these two networks share some of their backbone links to verify the information we obtained from each one separately.\footnote{Even for identical pairs of PoPs, there are small variations in the measured optical latencies in the ESnet data for 10 links, possibly due to optical switching delays, with an average difference of $31 \mu$s. With $5$ $\mu$s roughly corresponding to $1$ km of fiber, different measurements will cause a difference in link length estimates from a few to ~10 km. For each of these links, we used the minimum signal latency and the resulting length estimate as the true link length.} The link lengths in the two maps for the common backbone links are very close to each other\footnote{Except for two links, the first one is $150$ km longer in ESnet data, and the second one is $90$ km longer in Internet2 data. It is likely that these two links are inside separate conduits following different routes.}, so most likely that link lengths in Internet2 data also come from optical signal latency tests.

\begin{figure}[t]
	\centering
	\includegraphics[width=3.3in]{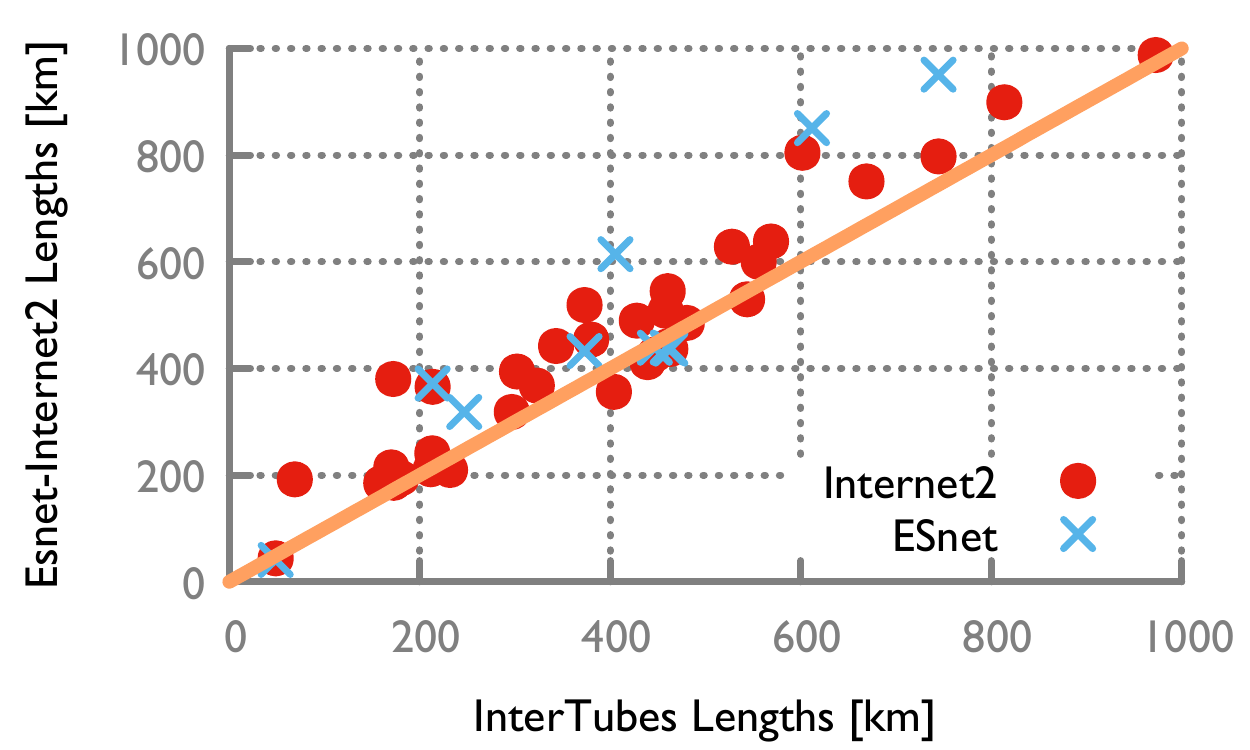}
	\caption{Comparing the lengths of common links between InterTubes and Esnet and Internet2 backbones}
	\label{fig:esnet-internet2}%
\end{figure}

We examined how close the fiber-optic link lengths obtained from ESnet and Internet2 are to the LOS distances between the link endpoints. Figure~\ref{fig:length-los-ratio} plots the CDF of link length to LOS distance ratio for these two networks. The CDF for ESNet is clearly separated from the rest with median value $1.58$. The CDF for Internet2 also lies to the right of the CDFs for other networks, though it is closer to the curve for Akamai ICN, with medians $1.40$ for Internet2 and $1.34$ for ICN. The CDFs for Internet2 and ICN are close to each other compared to other networks, and this is somewhat explained by the physical maps of the two networks being strikingly similar to each other, with many common endpoints and long-haul links between same pairs of cities.

In both ESnet and Internet2 backbones, there are some common links with the InterTubes dataset, and figure~\ref{fig:esnet-internet2} compares the lengths of common links in either Internet2 or ESnet with the InterTubes dataset. All the common links but two, the lengths of which are plotted in the figure are listed in Level3's backbone in the InterTubes dataset. 
Since the link lengths in ESnet and Internet2 are based on signal latency tests, measured cable length will be longer than the conduit length as explained in section~\ref{sec:fiber-latency-detailed}. 
Still, we see some data points on or below the diagonal in the figure. However the differences in length for these are rather small and can be expected due to minor differences in endpoint locations or due to the lack of precise knowledge of fiber routes. The link with the largest difference in lengths is not in Level3's backbone according to the InterTubes dataset.

A large majority of the points appear above the diagonal as expected; for these points, lengths in ESnet/Internet2 are on average $30\%$ longer than the conduit lengths in InterTubes dataset, in the median the difference in lengths is $18\%$. 
These numbers are plausible given that DCF overhead and excess fiber inside the conduits will increase the cable length as discussed before. There are four data points for which the difference in lengths is more than $50\%$; this difference in cable length might be due to different fiber routes as the combined overhead of excess fiber length and DCF is typically smaller. Using a smaller number of links common in ESnet and Internet2 with length differences less than $1\%$, we obtained the shortest fiber route lengths from Zayo's backbone between the endpoint (PoP) locations in ESnet. For all links, cable length in ESnet is found to be larger than the conduit length we computed, though the differences lie in a wide range between $5\%$ and $49\%$, with an average difference of $20\%$. 


\subsection{Latency in the ESnet Backbone}
Both Internet2 and ESnet constantly measure the throughput, loss and latency between their PoPs, and some of this data is available for researchers~\cite{esnet-data,internet2-data}. 
We downloaded traceroutes performed between all directly connected major POPs in ESnet backbone~\cite{esnet-backbone}; in total we have traceroutes in both forward and reverse directions for $24$ long-haul links. Our data covers a $3.5$ month period between January 1 and April 18, 2017. Traceroutes were performed every 10 minutes. On average we have $\sim13.8$K traceroutes for each link and the minimum number of traceroutes for any link is $\sim8.6$K. 

Each hop on the traceroutes has names identifying its location, and we have not observed any traceroutes which indicate a longer path was taken instead of the target fiber link. The fact that intended direct fiber links are used is shown by the very closely matching $c$-latency and observed RTTs. Table~\ref{table:esnet-inflation} shows the summary of the results, where we show the inflation over all the links both for minimum and average RTT observed for each link during the measurement period. We see that even the average RTTs are much closer to $c$-latency compared to minimum RTTs in the CDN and AT\&T data. However this is not very surprising, since we have very accurate fiber link lengths for ESnet as explained previously (in section~\ref{sec:length-compare}) and the shortest physical routes through the direct fiber links are taken. 
The largest inflation is observed for the shortest link we examined
(between Sacramento and Sunnyvale, $244$ km), and the inflation in minimum latency in one direction was $22\%$. However, the difference between minimum RTT and $c$-latency is only $0.3$ ms in one direction and $0.5$ ms in the other, which is small.

\begin{table}
	\centering
	\begin{tabular}{lccc}
		& \textbf{Min.} & \textbf{Median} & \textbf{Max.} \\
		\hline
		\textbf{Min. RTT}	& 1.010	& 1.031	& 1.215 \\
		\textbf{Avg. RTT} & 1.014	& 1.048	& 1.250
	\end{tabular}
	\caption{Latency inflation  in ESnet backbone over directly connected POPs}
	\label{table:esnet-inflation}
\end{table}
\subsection{Latency in the Internet2 Backbone}
Using the available data for researchers at Internet2 measurement archives, we obtained latencies between all pairs of $11$ major Internet2 PopS, which are the vantage points underlying the Internet2 performance dashboard~\cite{internet2-data}.
Latencies we have are obtained using One Way Active Measurement Protocol (OWAMP)~\cite{owamp} which relies on synchronized clocks by NTP. We obtained one-way measurements between $107$ pairs, so we have two-way latency data between $53$ pairs of hosts. Our data covers a one week period between April 12,2017 and April,19 2017. For each pair of PoPs, we have measurements every minute of the day for the entire week. To detect even tiny amounts of packet loss, one-way delay measurements are performed every $100$ ms, so we have statistics of $600$ one-way delay measurements for every minute.

\begin{figure}[t]
	\centering
	\includegraphics[width=3.3in]{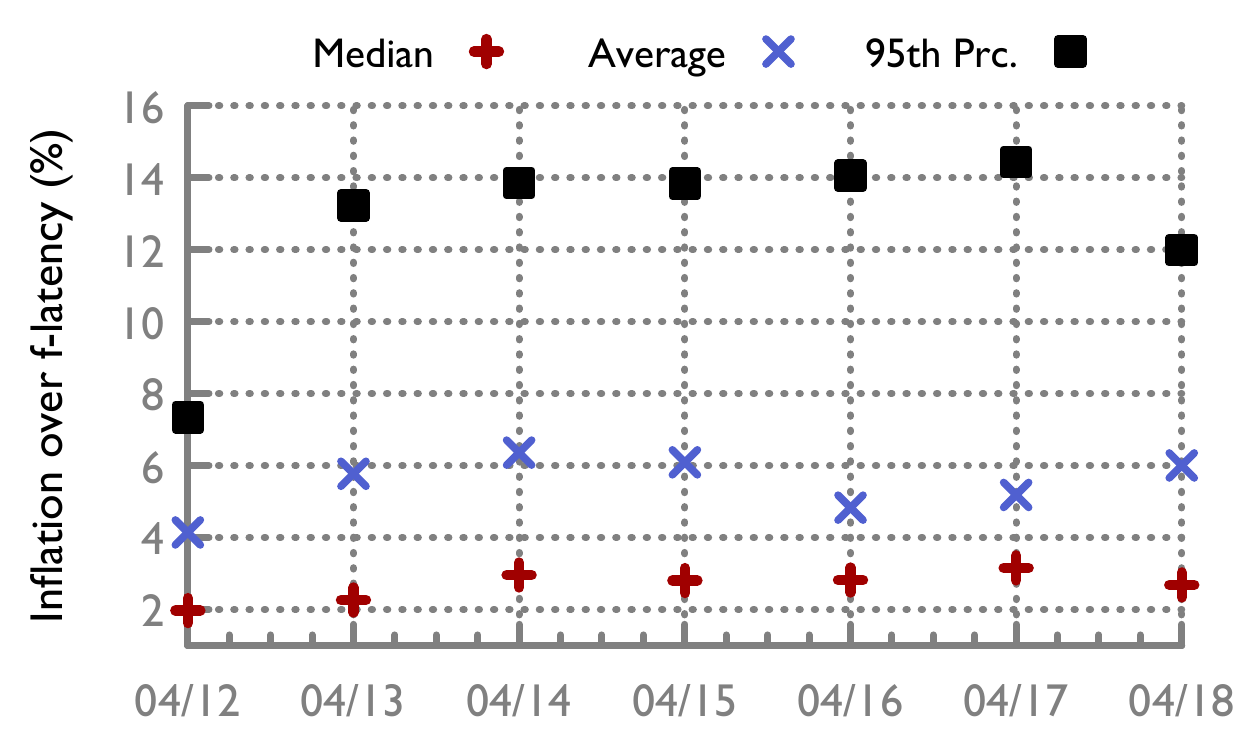}
	\caption{Change of inflation in min $1$-way delay over $f$-latency in Internet2 during one week in April 2017.}
	\label{fig:internet2-latency}%
\end{figure}

For each city pair, we computed the minimum one-way delay for each minute, for every day. Then, we averaged the minimum one-way delays in two directions for each of $53$ pairs of cities so that any effect of clock skew is cancelled. Using the Internet2 topology information and link lengths, we computed the length of the shortest path between each pair of PoPs, and computed the inflation of averaged one-way delays over $f$-latency. Over the one week period, we observed that latencies are consistently very low and very close to $f$-latency. For example, for April 12, 2017 the pair of cities with minimum inflation is just $0.02\%$ inflated over $f$-latency, whereas the median, average and $95$-th percentile of inflation over $f$-latency (over all city pairs) are found as $1.97\%$, $4.14\%$ and $7.35\%$ respectively. Figure~\ref{fig:internet2-latency} shows how these statistics change during the measurement period. 
Minimum inflation over the week stays always very close to $0$, and the median inflation shows little change between $1.97\%$ and $3.15\%$. Similarly, mean inflation shows little change between $4.14\%$ and $6.36\%$. As expected, $95$-th percentile inflation shows larger deviations and changes between $7.97\%$ and $14\%$. Overall, we see that the vantage points are well distributed in USA and the distances, and the shortest path lengths between them are very long, causing the packets to pass through multiple transport elements for most of them. Despite this, we observe that the measured delays are very close to $f$-latency. Note that the link lengths in Internet2 already capture the latency overhead in the optical layer along fiber conduits, and the overhead of due to traversing multiple conduits and optical switching seem small. 

\section{Related Work}
\label{sec:related}

\paragraphb{Fiber backbones:}
Estes and Hart~\cite{coxBackbone} explain how the Cox national fiber backbone was built using $12,000$ miles of owned and leased dark fiber. Digital ROADMs were used to engineer the network, enabling quick rerouting around link failures and improving availability. The authors also discussed how quality of the fiber was an important consideration to avoid extra signal regeneration and reduce the impact of dispersion compensation, leading to choosing a commercial G.655 NZ-DSF fiber for $85\%$ of the links. Doverspike et al.\ discuss the structure of ISP backbones, including the fiber, DWDM and ROADM layers, and show both the complexity and evolution of the technologies used in the underlying physical layers~\cite{ispStructure}. Das et al.\ advocate for redesigning IP core networks to cut costs by replacing expensive backbone routers with hybrid packet-optical switches~\cite{rethinkingCore} and their work includes an overview of typical ISP backbones. Ramamurthy et al.\ also advocate for a redesign of backbone networks to cut costs by removing redundant routers by leveraging the capabilities of the optical transport network~\cite{ip-over-wdm}. Their work also includes a nice overview of ISP backbone architectures. 

\paragraphb{Optical layer:}
Recent works from Microsoft researchers and academics focus on improving the availability and capacity of fiber-optic links.
Filer et al.\ focus on making better use of the fiber in the Microsoft backbone to improve network capacity by using Bandwidth Variable Transceivers~\cite{elasticOptical}. On a $4000$ km testbed link built using fiber and equipment disassembled from links in their backbone, they showed that the capacity of existing links can be improved as much as $70\%$ with the addition of elastic transceivers.
Ghobadi and Mahajan analyzed the durations and characteristics of optical layer outages in the Microsoft backbone~\cite{opticalFailures}. Their work shows that observing the drops in signal quality can be used to predict outages, and  the authors argue that WAN traffic engineering and control planes should incorporate information from the physical layer. Singh et al.\ follow up on these efforts, analyzing the SNR of $2000$ WAN links in MS backbone and arguing for dynamically adjusting link capacities based on SNR instead of declaring the links with SNR below a certain threshold down~\cite{dynamic-bw-fiber}. The challenge of increased complexity of Traffic Engineering due to variable link capacities is addressed by proposing a graph abstraction with an augmented topology which enables using existing Traffic Engineering algorithms without change.

\paragraphb{Reducing latency in the wide area:}
Earlier academic and industry efforts showed the benefits of using overlays to reduce latency and improve availability~\cite{RON-Andersen,AkaOverlay-Ramesh1,AkaOverlay-Ramesh2}. Other works also proposed using CDN redirections to find low-latency paths~\cite{akamai-overlay}.
In ~\cite{stitchingIXP}, Kotronis et al.\ put forward a framework for stiching inter-domain paths at IXPs. This stitching can be performed under a variety of objectives including minimization of latency. This technique can overcome latency inflation imposed by inefficient inter-domain routing. Cloud-Routed
Overlay Networks (CRONets) ~\cite{cronets} rely on nodes from IBM and Amazon cloud for the same purpose. While the focus of this paper was on bandwidth, the ideas can be easily extended to latency.

\paragraphb{Measurement studies:}
Latency inflation in inter-domain paths is analyzed in ~\cite{Spring:2003}. A recent work examines the path inflation for mobile traffic~\cite{zarifis}. Zhu et al.\ examine the causes of latency increases between the servers of a CDN and find that $40\%$ of the increases are caused by BGP route changes~\cite{latlongGoogle}. Our focus in this has been finding the causes of latency overhead observed over direct fiber-optic links. 
A small scale study of latency inflation in optical networks is performed  in~\cite{Noutsios:04}.
The authors had accurate link lengths and detailed knowledge of the existing network elements along the links they measured including Dispersion Compensating Modules (DCM) adding $15\%$ latency overhead. Despite the detailed knowledge of optical links and transport equipment, measured and calculated RTTs differed by $9\%$ for some links.
\section{Discussion}
\label{sec:discussion}
When we initiated the measurements from the CDN servers near the fiber conduit endpoints, we were naively expecting to observe RTTs close to $f$-latency based on conduit lengths. With latency data from multiple sources revealing a more complex picture, what is a reasonable expectation of RTT between two locations if we approximately know the conduit length? Surprisingly this is not an easy question to answer. Even comparing the link (cable) lengths in ESnet and Internet2 to conduit lengths in the InterTubes dataset provided limited help since observed differences in length lie in a wide range. We know that with modern NZ-DSF fibers DCF overhead will be around $5\%$, but if the fiber type is older, classic SMF it can add as much as $20\%$ to the latency. We should expect at least $5\%$ overhead due to fiber slack left during construction, but the excess cable length can be even higher either due to  tube/cable design or purposeful additions for price differentiation. Attempting to answer the initial question, we believe RTTs over at least $20-25\%$ of $f$-latency based on conduit length should be expected in most cases unless we have a route optimized for low latency. 

When we only know the distance between two places, we also need to take the circuitousness of the fiber routes into account before estimating latency. InterTubes data shows conduit lengths are just $20\%$ longer than LOS distances in the median, but the conduit lengths we obtained from ICN ($34\%$) and Zayo data ($27\%$) exhibit higher ratios. On top of this, we have the overhead due to extra cable length and the overhead in the optical layers as described above.
For Internet2 and ESnet, $40\%$ and $57\%$ inflation in the median capture both the overhead in the physical layer and the circuitousness of the paths. In light of these numbers, Google's mentioning $50\%$ latency inflation over geodesic distance indicating a near-identical configuration~\cite{googleRuleOfThumb} seems reasonable, although it might be optimistic in some cases.

Despite the mentioned overheads, we believe the large fiber footprint and availability of multiple routes between many locations show a great opportunity for building a network as analyzed in section~\ref{sec:fiberDesign}, even though it will require careful examination of fiber routes from different fiber providers in addition to the underlying technology. Note that in addition to leasing short term wavelengths, it is also possible to lease dark fiber over selected routes and update the transport equipment to reduce latency overhead even on older fiber installations.

\section{Acknowledgements}

This research used data from ESnet, which is supported by the Office of Science of the U.S. Department of Energy under contract DE-AC02-05CH11231.

\bibliographystyle{abbrv}
{
\bibliography{paper}
}

\end{document}